\documentclass[14pt,aps,preprint]{revtex4-1}
\usepackage[utf8]{inputenc}
\usepackage[T1]{fontenc}
\usepackage{amsmath}
\usepackage{amsfonts}
\usepackage{amssymb}
\usepackage{graphicx}
\usepackage{lineno}
\usepackage{graphicx}
\usepackage{amsmath}
\usepackage{bbold}
\usepackage{amssymb}
\usepackage{color}
\usepackage[normalem]{ulem}
\usepackage{epsfig}
\usepackage{amssymb,amsfonts,amsmath,color}
\usepackage{array}
\usepackage{graphicx} % Required for inserting images
\usepackage{float}
\begin{document}
\title{Asymmetric Variability: The Impact of Uneven Stochasticity on Competitive Dynamics}
\author{Ori Turkia and Nadav M. Shnerb}
\affiliation{Department of Physics, Bar-Ilan University,
Ramat-Gan IL52900, Israel.}
\date{September 2024}

\begin{abstract}
    \noindent Competition between species and genotypes is a dominant factor in a variety of ecological and evolutionary processes. Biological dynamics are typically highly stochastic, and therefore, analyzing a competitive system requires accounting for the random nature of birth and death processes (demographic stochasticity) as well as the variability of external conditions (environmental stochasticity). Recent studies have highlighted the importance of species life history, showing that differences in life history lead competing species to experience different levels of demographic stochasticity. Here, we propose a simple model of two-species competition with different life histories and derive analytical expressions for various properties (fixation probability, fixation time, absorption time, probability density) under a wide range of conditions, including migration, selection, and environmental stochasticity. These properties provide insights into the long-term outcomes of competition, such as species persistence, extinction risks, and the influence of environmental variability on the evenness of the community.
\end{abstract}

\maketitle

\section{introduction}

Understanding species interactions and  species  turnover is essential for addressing pressing ecological challenges. A clearer grasp of these dynamics can provide valuable insights into mitigating the alarming loss of biodiversity observed in numerous ecosystems~\cite{cardinale2012biodiversity,dornelas2014assemblage} - a problem with profound global consequences. Moreover, analyzing species interactions plays a pivotal role in efforts to modify or optimize microbial communities, such as the human microbiome~\cite{fassarella2021gut} or soil microbiome~\cite{zhang2021theory} - areas that have become focal points of extensive research. Additionally, our theoretical understanding of evolutionary dynamics hinges on accurately describing ecological interactions, as competitive pressure is the central driving forces of evolution.

A central challenge in ecology is identifying the conditions under which competing species can coexist. Despite significant progress, this question remains a fundamental puzzle, as the competitive exclusion principle~\cite{hardin1960competitive,tilman1982resource} predicts that one species—typically the stronger competitor—will eventually outcompete and exclude the other. Within this context, stochasticity plays a crucial role in shaping species coexistence~\cite{chesson2000mechanisms}. Biological and ecological systems are inherently stochastic, with random fluctuations capable of driving diverse outcomes. Under certain conditions, such fluctuations can even facilitate coexistence~\cite{chesson1981environmental,barabas2018chesson}. Understanding these stochastic dynamics is therefore essential for building robust ecological theories and addressing the related challenges.

Stochasticity in ecological systems is typically categorized into two main types: demographic stochasticity and environmental stochasticity~\cite{lande2003stochastic}. Demographic stochasticity arises from random variation in the reproductive success of individuals. For instance, each individual might have an equal probability of producing either zero or two offspring. In contrast, environmental stochasticity affects entire populations in a correlated manner. For example, during favorable years, each individual may, on average, produces two offspring, whereas in unfavorable years, this average might drop to just one-half.

Neutral models, where stochasticity is the sole driver of abundance variations, have garnered significant interest over the years. In their simplest form~\cite{kimura1985neutral,ewens2012mathematical,karlin1981second,azaele2015towards}, these models assume that competing species are competitively identical, i.e., that any differences between them are irrelevant to the reproductive process. Both species compete for the same resources, and the stochasticity affecting them is purely demographic.  Competitive forces manifest themselves through a constraint on the total population size. In this framework, the model is considered neutral if individuals from both species share the same offspring distribution or, at the very least, distributions with identical means and variances.

Neutral models of this kind, whether describing competition between two biological types or communities composed of many species, are well established in the fields of population genetics and ecological community dynamics~\cite{hubbell_book,maritan1}. These models have been successful in explaining observed patterns, such as species abundance distributions. However, they often fall short in capturing temporal fluctuations in species abundances and related metrics~\cite{leigh2007neutral,kalyuzhny2014temporal,kalyuzhny2014niche,chisholm2014temporal}.

To address this limitation, time-averaged neutral models have been proposed over the past decade~\cite{kalyuzhny2015neutral,danino2018theory,dean2020stochasticity,van2024tiny}. In these models, environmental conditions can temporarily favor one species over others, granting it a competitive advantage at a given moment. However, when averaged over time, the long-term fitness of all species remains equal, preserving the core assumption of neutrality.

In a recent study, \citet{jops2023life} proposed a novel type of neutral model. These models exclude environmental stochasticity, and the average number of offspring per individual is identical for both species. However, the species differ in their life history traits, differences that can be empirically quantified using a population matrix model. These traits result in varying magnitudes of demographic stochasticity between species. For example, an individual of one species might produce four, five, or six offspring with equal probability (yielding an average of five), while an individual of the competing species could produce anywhere between two and eight offspring with equal probability. Although the averages are identical, the intensity of demographic stochasticity differs, leading to distinct patterns of population fluctuations.

In this work, we present a simple model of competition between two species with different life histories. Using this model, we analyze key aspects of interspecific competition when the competing species experience differing levels of stochasticity:

\begin{enumerate}
    \item In the simplest case, where stochasticity is purely demographic and dynamics occur on an isolated island, we employ the diffusion approximation to derive expressions for the probability of ultimate fixation, the time to absorption, and the time to fixation (Section~\ref{sec4}).

    \item When the system is subject to constant immigration of individuals, rendering the extinction state non-absorbing, we derive formulas for the equilibrium distribution (Section~\ref{sec5}).

    \item We examine the interplay between stochasticity and selection, with a particular focus on the scenario where the species that is {\it stochastically superior} - experiencing lower stochasticity - is also {\it selectively inferior}, i.e., having a lower mean number of offspring per individual (Section~\ref{sec6}).

    \item We also explore the relationship between demographic and environmental stochasticity. Once again, the most intriguing scenario arises when the species advantaged by demographic stochasticity is disadvantaged by environmental stochasticity (Section~\ref{sec7}).

\end{enumerate}

In the penultimate section, we aim to provide a broader perspective, clarifying how, why, and when stochasticity undermines a species' effective fitness in competition. Finally, we present a general discussion of the conclusions emerging from our analysis.

\section{The model}

We model the competition between two species with non-overlapping generations, each characterized by a distinct life history. In its simplest form, the dynamics proceed through four sequential steps:

\begin{enumerate}

\item {\bf Seed Production}: Adult individuals of each species produce seeds, with the quantity determined by their reproductive strategies and environmental conditions.

\item {\bf Adult Mortality}: At the end of each generation, all adult individuals die, ensuring complete generational turnover and the absence of overlap between successive generations.

\item {\bf Juvenile Recruitment}: Seeds germinate and recruit into a juvenile pool, resulting in a total of  $J$ juveniles.

\item {\bf Maturation and Seed Production}: Each juvenile has a species-specific probability of maturing into an adult capable of producing seeds, thus forming the next generation.

\end{enumerate}

We will analyze the transition from one juvenile stage to the next. At a given stage, if the focal species is represented by $n$ juveniles, its fraction is $x = n / J$. The rival species then consists of $J - n$ juveniles, corresponding to a fraction of $1 - x$.

The life-history difference between the two species is expressed in two parameters: the probability of a juvenile reaching maturity and the number of seeds produced by each adult. In particular, the probability of each juvenile to mature and produce seeds is $p_f$ for a focal species individual and $p_r$ for a rival species individual. Correspondingly, an adult of the focal species produces, on average, $m/p_f$ seeds, while a rival species adult produces $m/p_r$ seeds. Therefore, the mean number of seeds per individual is the same, $m$, for both species, and the variance is species-dependent.

This model can be viewed as a neutral model, where no species has a deterministic selection advantage since the average number of offspring is identical. However, differences in the level of stochasticity experienced by each species provide an advantage, as we shall see, to the more stable species.

Given that, the total number of seeds in the soil (seed bank) at the end of step 1 is,

\begin{equation} S = S_f + S_r = m \ell_f / p_f + m \ell_r / p_r, \end{equation}

where $\ell_{f}$ ($\ell_{r}$) represents the number of reproducing individuals of the focal (rival) species. These numbers are drawn from a binomial distribution with $n$ ($J - n$) trials and a success probability of $p_f$ ($p_r$).

In the recruitment step, seeds are randomly selected from the seed bank to fill each of the $J$ available slots (assuming no spatial effects and a well-mixed dynamic). Therefore, the number of focal species juveniles in the next generation, $n_{t+1}$, follows a hypergeometric distribution with $m \ell_f / p_f$ favorable objects out of $S$.

When $m$ is large, the no-replacement constraint becomes negligible. Consequently, $n_{t+1}$ can be approximated by a binomial distribution with $J$ trials and a success probability given by
\begin{equation}
Q = \frac{\ell_f / p_f}{\ell_f / p_f + \ell_r / p_r}. \end{equation}

The dynamics (1-4), as described so far, do not support a stable coexistence state. Starting from any initial condition, the long-term outcome will inevitably lead to the fixation of one species. In such a scenario, the primary questions of interest are:

\begin{itemize}
   \item What is the chance of ultimate fixation, $\Pi(x)$, of the focal species, given its initial fraction (at the juvenile stage) $x$.

\item What is the mean time to absorption, $T_A(x)$, i.e., the mean time until one of the species goes extinct.

\item What is the mean time to fixation, $T_F(x)$, the mean time until the focal species reaches fixation, conditioned on fixation.
\end{itemize}

\section{The diffusion approximation: mean and variance of $\Delta x$}

In what follows, we implement the diffusion equation and its related techniques, such as the backward Kolmogorov equation. These techniques rely on the mean and variance of abundance variations, which we calculate below.

We approximate $\ell_f$ and $\ell_r$ using the mean plus-minus the standard deviation:

\begin{equation}
\label{eq3} \ell_f = p_f n \pm \sqrt{n p_f (1 - p_f)}, \qquad \ell_r = p_r (J - n) \pm \sqrt{(J - n) p_r (1 - p_r)}.
\end{equation}
Correspondingly, the total number of seeds is given by:
\begin{equation}
\label{eq3} S_f^{\pm} = m n \pm m \sqrt{n (1 - p_f) / p_f}, \qquad S_r^{\pm} = m (J - n) \pm m \sqrt{(J - n) (1 - p_r) / p_r}.
\end{equation}

Thus, we have four scenarios ($++, +-,-+,--$), each with probability $1/4$. The fraction of focal species seeds in the seed bank is $$ Q^{++} = S_f^+/(S_f^+ + S_r^+), \ Q^{+-} = S_f^+/(S_f^+ + S_r^-),$$  $$ Q^{-+} = S_f^-/(S_f^- + S_r^+), \ Q^{--} = S_f^-/(S_f^- + S_r^-). $$

Now we have to decide about the procedure of drawing:

{\bf Drawing seeds with replacement.}
In the case where $m \gg 1$, we can implement the binomial distribution in the seed recruitment step. The probability of the focal species capturing a slot is given by the appropriate value of $Q$. The mean and the variance, given $Q$, are therefore:

\begin{equation} {\rm mean}[n_{t+1}] = JQ, \qquad {\rm Var}[n_{t+1}] = JQ(1-Q). \end{equation}

By taking the mean and the variance over all four scenarios, each with probability $1/4$, one finds, to the leading order in $1/J$:

\begin{equation} \label{mean}
\mu =  \overline {\Delta x} = \overline{(n_{t+1}-n_t)}/J = \frac{x(1-x)(p_f-p_r)}{Jp_fp_r}.
\end{equation}
and
\begin{equation} \label{variance}
{\rm Var}[\Delta x] = \frac{x(1-x)[p_r(1-x)+p_f x]}{Jp_f p_r}.
\end{equation}

\begin{figure}[h]
	%\vspace{-3.cm}
	\centering{
		 \includegraphics[width=6cm]{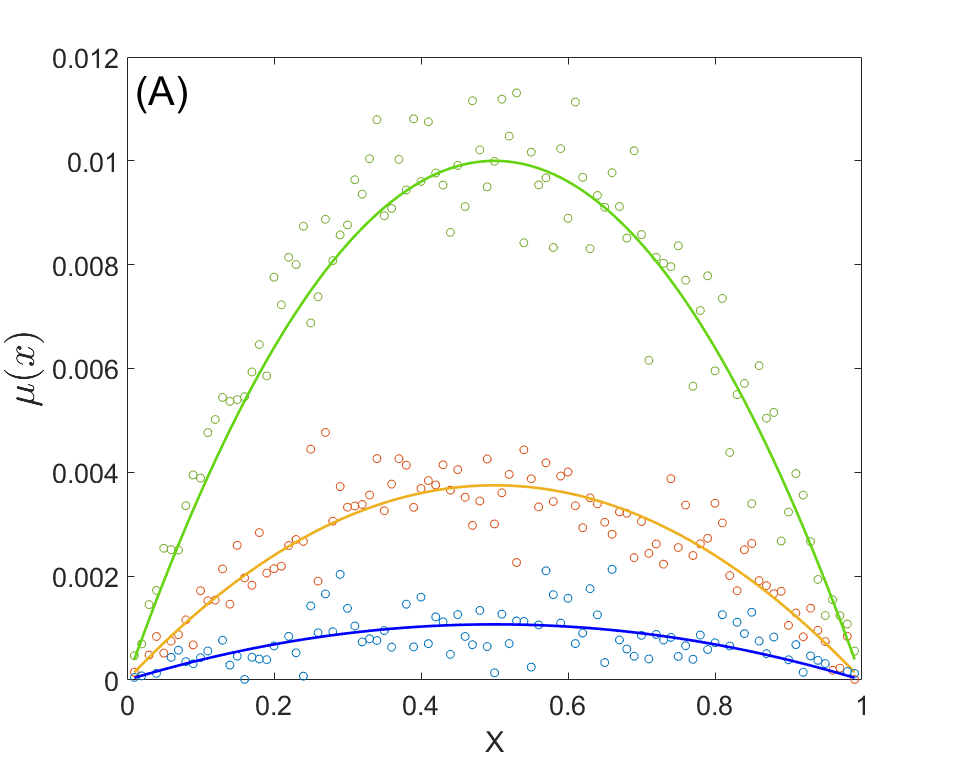} \includegraphics[width=6cm]{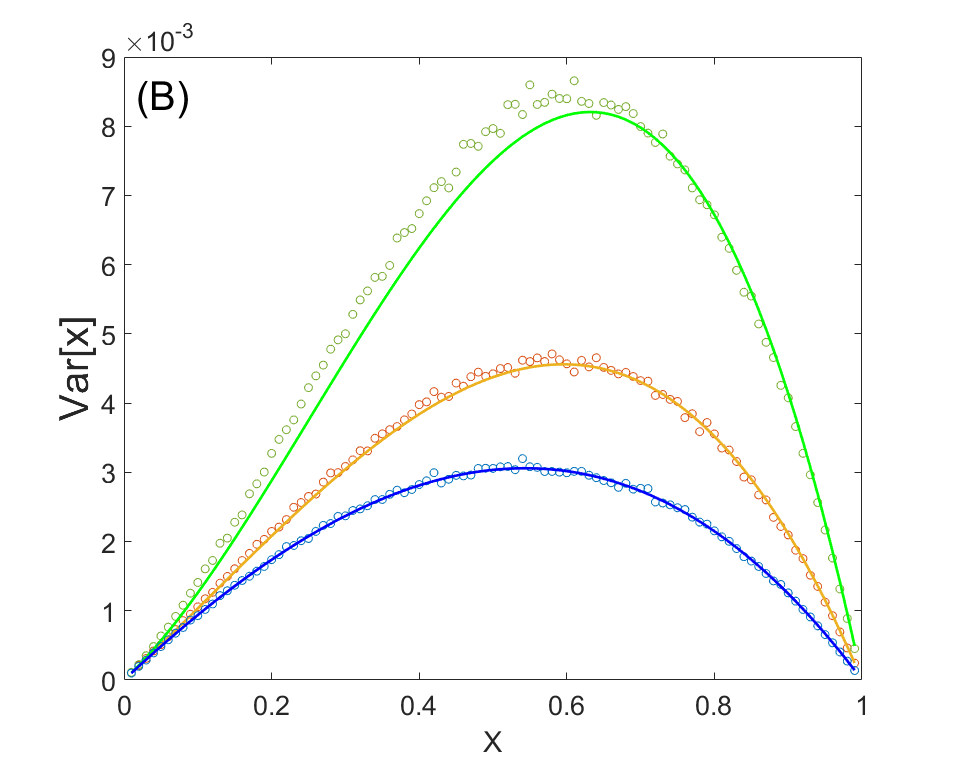}}
	%\vspace{-3.cm}
	\caption{{\bf Mean (panel A) and variance (panel B) of $\Delta x$} are plotted for $p_r=0.2$ (green), $p_r=0.4$ (orange), and $p_r=0.7$ (blue) in a system with $J=100$ individuals. For each $1 < n < 99$, $\Delta x = (n_{t+1} - n_t)/J$ was calculated $10^4$ times to estimate these quantities. The solid lines represent the theoretical predictions given by Eqs. (\ref{mean}) and (\ref{variance}). In all cases, $p_f = 1$.   \label{fig1}}
\end{figure}

Figure \ref{fig1} demonstrates the agreement between Eqs. (\ref{mean}) and (\ref{variance}) and the results obtained from simulations.

{\bf Drawing seeds without replacement.} When $m$ is not large enough, it is important to account for the fact that, as seeds are drawn from the seed bank, the total number of seeds decreases. Mathematically, this means that the probability of selecting $n_{t+1}$ focal species seeds follows a hypergeometric distribution, representing the chance of obtaining $n_{t+1}$ successes in $J$ draws from a seed population of size $S$ containing $Q$ focal species seeds.

In this case (again, in the limit $J \gg 1$), the mean remains the same as in Eq.(\ref{mean}), but the variance is given by:
\begin{equation} \label{var2}
{\rm Var}[\Delta x]= \frac{x(1-x)(p_r m (1-x)+p_f m x-p_f p_r)}{J m p_f p_r}.
\end{equation}

\begin{figure}[h!]
	%\vspace{-3.cm}
	\centering{
		 \includegraphics[width=8cm]{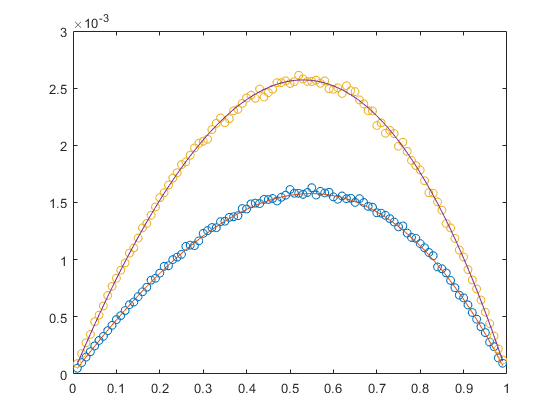}
	%\vspace{-3.cm}
	\caption{{\bf Variance of $\Delta x$ at the no-replacement case} for $J=100$, $p_r = 0.8$, $p_f =1$ comparing $m=2$ (blue) and $m=10$ (yellow). The full lines are the corresponding analytic predictions, Eq. (\ref{var2}).  }  \label{fig2}}
\end{figure}

Note that (\ref{var2}) converges to (\ref{variance}) as $m \to \infty$. Figure \ref{fig2} demonstrates the agreement between the expression (\ref{var2}) and the numerical results for the non-replacement procedure with various $m$-s.

\section{Chance of fixation, time to absorption and time to fixation} \label{sec4}

Once the mean and the variance are given, one can calculate the chance of ultimate fixation, $\Pi(x)$. This quantity satisfies,
\begin{equation}
  \frac{{\rm Var(x)}}{2}  \Pi''(x)+ \mu(x) \Pi'(x) = 0, \quad \Pi(0) = 0, \quad \Pi(1) = 1.
\end{equation}

\begin{figure}[h]
	%\vspace{-3.cm}
	\centering{
		 \includegraphics[width=5.cm]{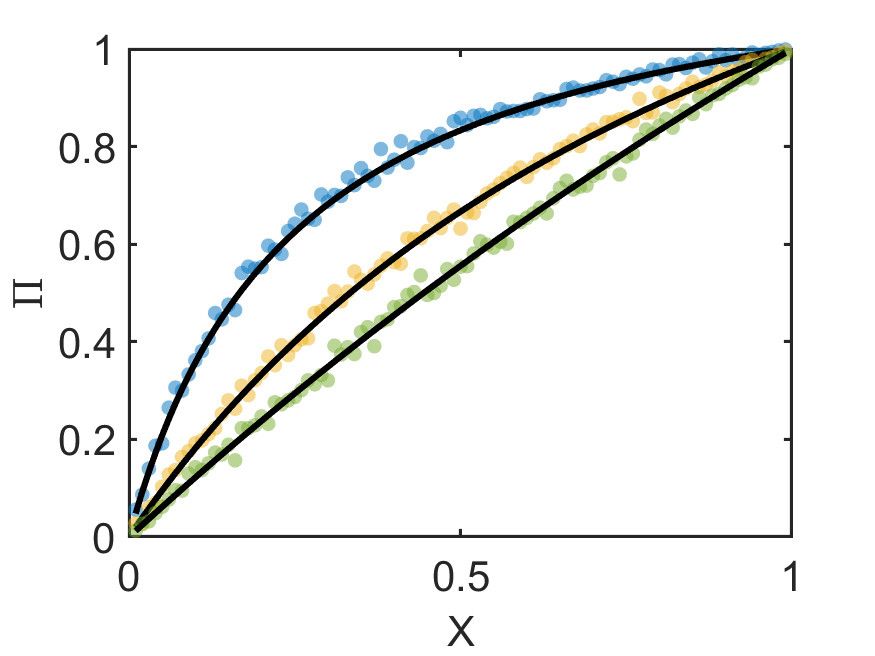} \includegraphics[width=5.cm]{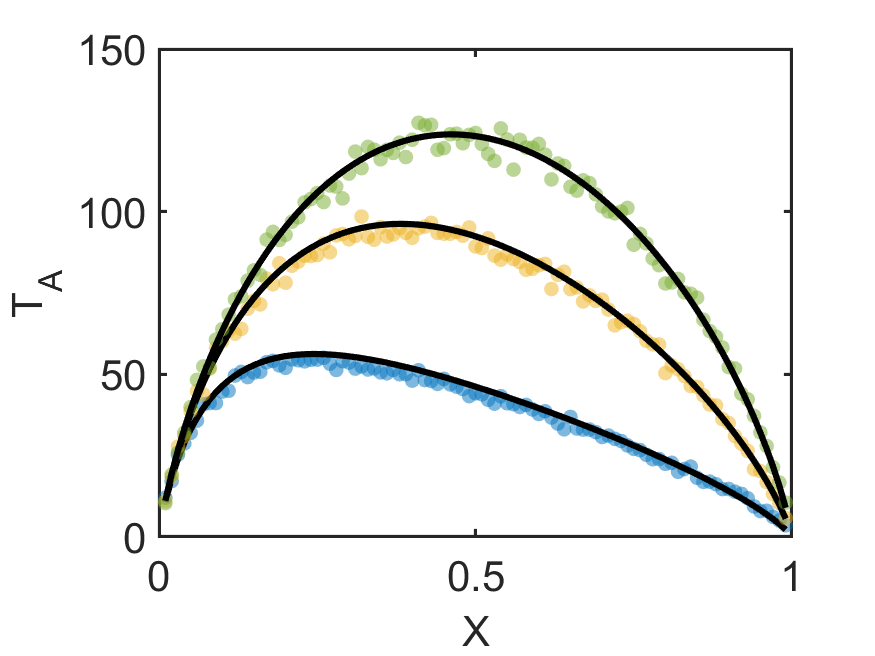} \includegraphics[width=5.cm]{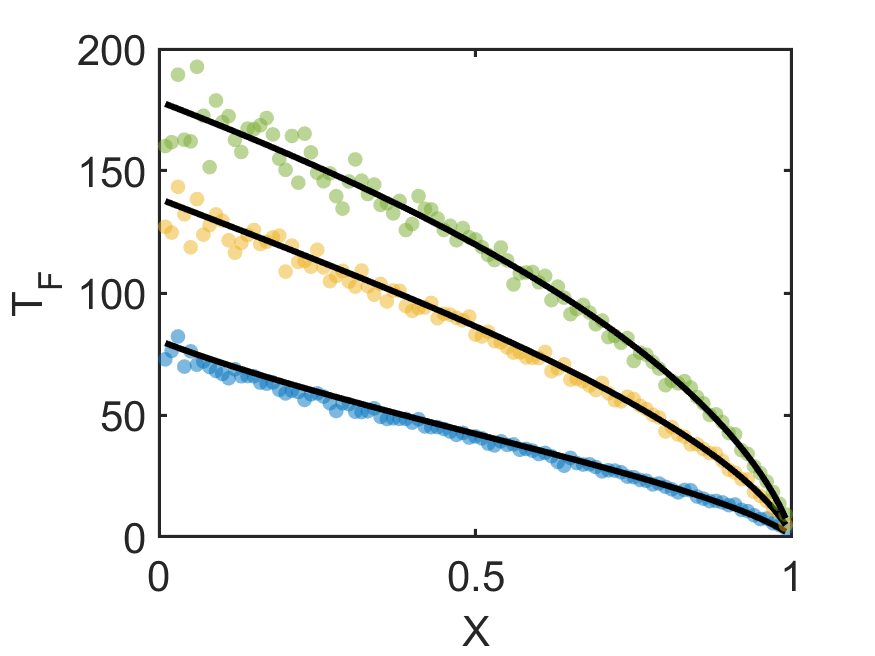}
	%\vspace{-3.cm}
	\caption{{\bf $\pi$ (left), $T_A$ (middle)  and $T_F$ (right) as a function of x}. Juveniles are recruited from seeds in the soil bank {\it with replacement}. Black lines depict  the analytic expressions, Eqs. (\ref{fix1}), (\ref{TA1}) and  (\ref{TF1}),  to be compared with the results of numerical simulations with $J=100$ and $m=5$ for  $p_r = 0.2$ (blue) $0.5$ (yellow) and $0.8$ (green). In all cases $p_f = 1$. Each circle represents an average  over $1000$ runs for $n \in [1..J-1]$.  }  \label{fig3}}
\end{figure}

In the with-replacement case, the result is~\cite{gillespie1974natural,jops2023life},
\begin{equation} \label{fix1}
    \Pi(x) = \frac{p_f x}{p_r(1-x) + p_f x}.
\end{equation}
The alignment of this result with the outcome of a numerical experiment is illustrated in Figure \ref{fig3}.

In case without replacement,
\begin{equation} \label{fix2}
 \Pi(x) = \frac{p_f x(m-p_r)}{mp_r(1-x) + m p_f x - p_f p_r }.
\end{equation}
as depicted in Fig. 4.

\begin{figure}[h]
	%\vspace{-3.cm}
	\centering{
		 \includegraphics[width=5.cm]{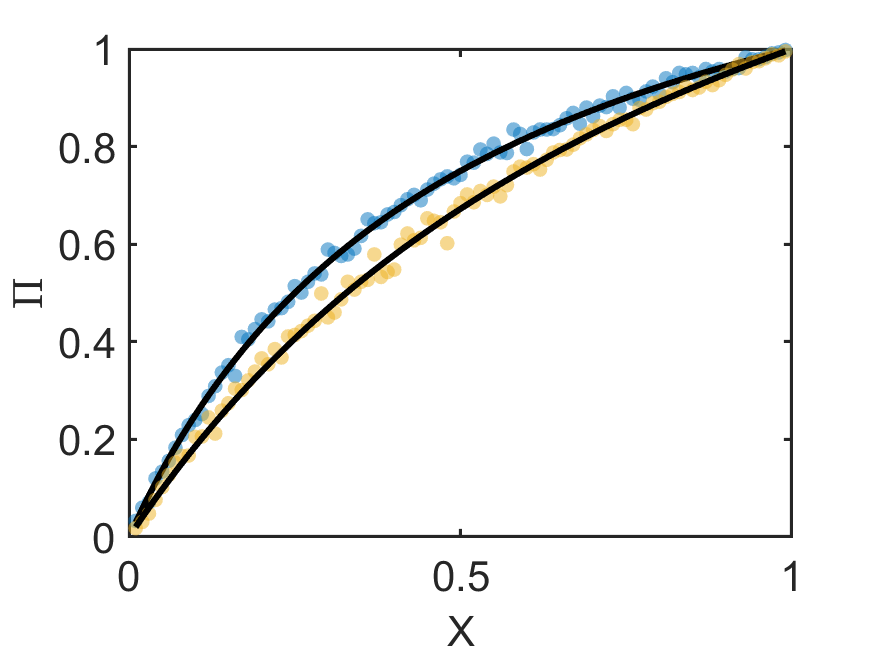} \includegraphics[width=5.cm]{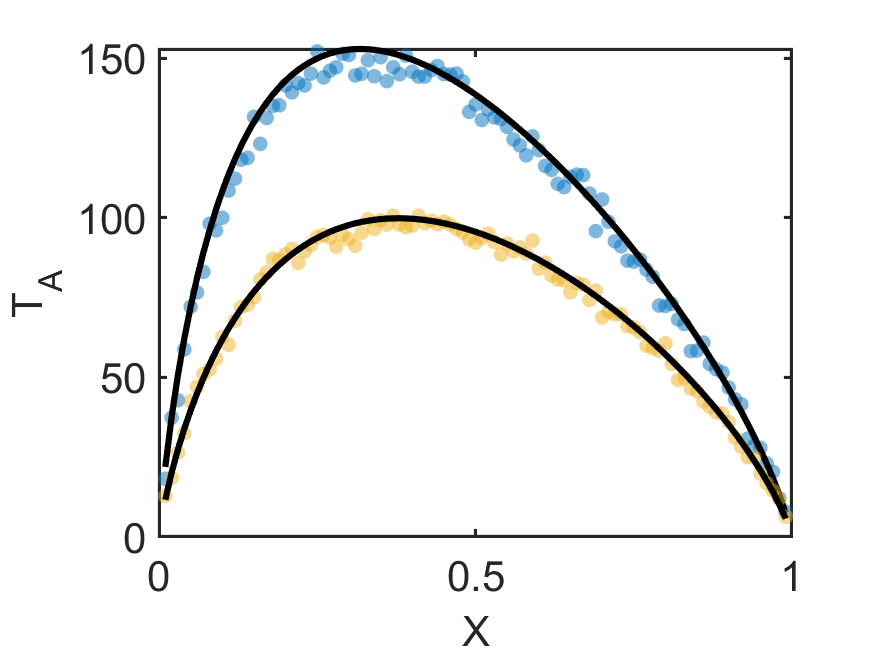} \includegraphics[width=5.cm]{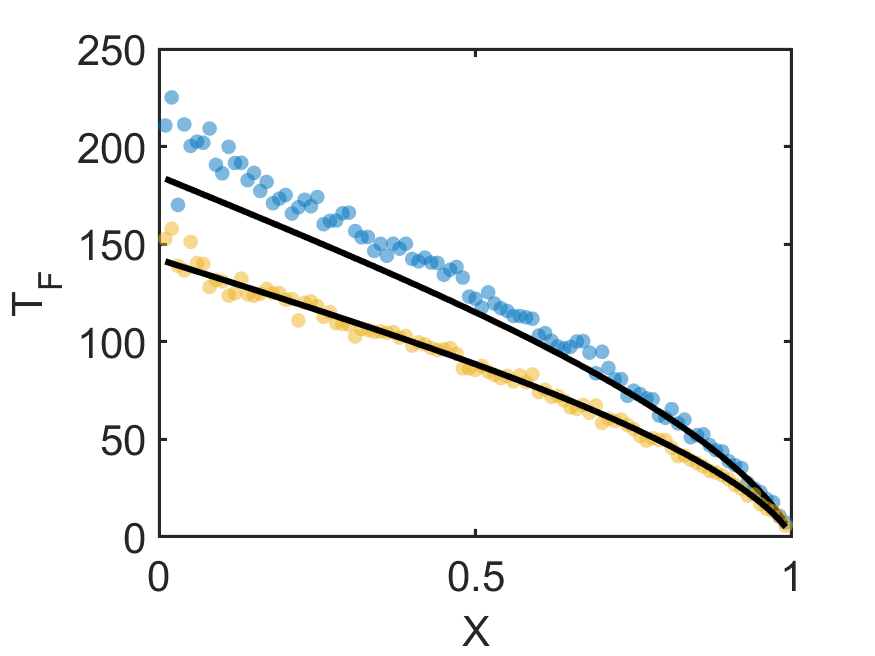}
	%\vspace{-3.cm}
	\caption{{\bf $\pi$ (left), $T_A$ (middle)  and $T_F$ (right) as a function of x}. Juveniles are recruited from seeds in the soil bank {\it without replacement}. Black lines depict  the analytic expressions, Eqs. (\ref{fix2}), (\ref{TA2}) and  (\ref{TF2}),  to be compared with the results of numerical simulations with $J=100$  for  $p_r = 0.5$. The case $m=2$  (blue) is compared, in each panel, with the case $m=20$ (yellow). In all cases $p_f = 1$. Each circle represents an average  over $1000$ runs for $n \in [1..J-1]$.  }  \label{fig4}}
\end{figure}

The time to absorption, $T_A$, satisfies,
\begin{equation}
  \frac{{\rm Var(x)}}{2}  T_A''(x)+ \mu(x) T_A'(x) = -1, \quad T_A(0) = 0, \quad T_A(1) = 1
\end{equation}

With replacement (middle panel of Fig. \ref{fig3}),
\begin{equation}  \label{TA1}
    T_A =-2J p_f p_r \frac{(1-x) \ln(1-x) + x\ln x}{p_r (1 - x) + p_f x}.
\end{equation}
This is Eq. (22) of \cite{jops2023life}.

Without replacement (middle panel of Fig. \ref{fig4}),
\begin{equation}  \label{TA2}
    T_A =-2Jm p_f p_r \frac{(1-x) \ln(1-x) + x\ln x}{m p_r (1 - x) + m p_f x-p_f p_r}
\end{equation}

Finally, to obtain the mean time to fixation, $T_F$, one calculates $Q$ according to,
\begin{equation}
  \frac{{\rm Var(x)}}{2}  Q''(x)+ \mu(x) Q'(x) = -\Pi(x), \quad Q(0) = 0, \quad Q(1) = 0
\end{equation}
and then
\begin{equation}
    T_F = Q(x)/\Pi(x).
\end{equation}

this yields, in case with replacement (right panel of Fig. \ref{fig3}),
\begin{equation} \label{TF1}
    T_F = 2 p_r J \frac{(p_r-p_f)(1-x)\ln(1-x)-[p_r(1-x)+p_f x]\ln[p_r (1-x)+p_f x]+(1-x)p_r\ln p_r +p_f x \ln p_f}{(p_f-p_r)x},
\end{equation}
and without replacement (right panel of Fig. \ref{fig4}),
\begin{equation} \label{TF2}
T_F = \frac{2  p_r J}{(m-p_r)(p_f-p_r)x} \left\{
 x C_2 \ln C_2 + (1-x) C_1 \ln C_1 -m(p_f -p_r) (1-x)\ln (1-x)
- f_1(x) \ln f_1(x)
 \right\},
\end{equation}
where
\begin{equation}
    f_1(x) =  m(p_r(1-x)+p_f x)-p_r p_f, \qquad C_1 = p_r (m-p_f) \qquad C_2 = p_f(m-p_r)
\end{equation}

\section{Equilibrium state under immigration} \label{sec5}

In this section, we analyze the stable state reached by the system, as described above, when competition takes place on an "island" with constant immigration of individuals, or seed dispersal, from the regional community ("mainland").

Let us first consider seed dispersal. We assume that the island receives $\theta$ seeds from each species annually, so the number of focal species seeds is $S_f = m \ell_f / p + \theta$, and correspondingly $S_r = m \ell_r / p + \theta$. The migration parameter in this context is $\nu = \theta / m J$, representing the ratio between the number of immigrant seeds and the mean number of local seeds.

To first order in $\nu$ and $1 / J$, seed dispersal does not affect the variance; however, it alters the mean, so Eq. (\ref{mean}) is replaced by:
\begin{equation}
\label{mean1} \mu = \overline {\Delta x} = \overline{(n_{t+1}-n_t)}/J = \frac{x(1-x)(p_f-p_r)}{Jp_fp_r}+\nu (1-2x).
\end{equation}

In a model with immigration (instead of seed dispersal), each juvenile, after the recruitment step, may be replaced by an immigrant from the mainland with probability $2\nu$. Since immigrants have an equal probability of being focal or rival, the net outcome remains unchanged. Similarly, a genetic model where the two 'species' (e.g., alleles) mutate into each other at a rate $\nu$ produces the same expression.

\begin{figure}[h]
	%\vspace{-3.cm}
	\centering{
\includegraphics[width=8cm]{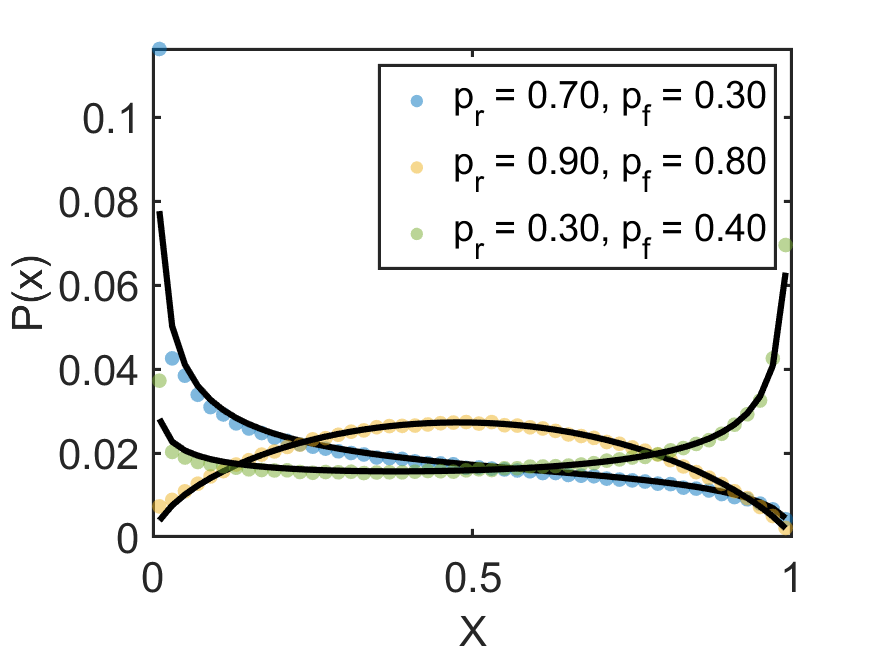} \includegraphics[width=8cm]{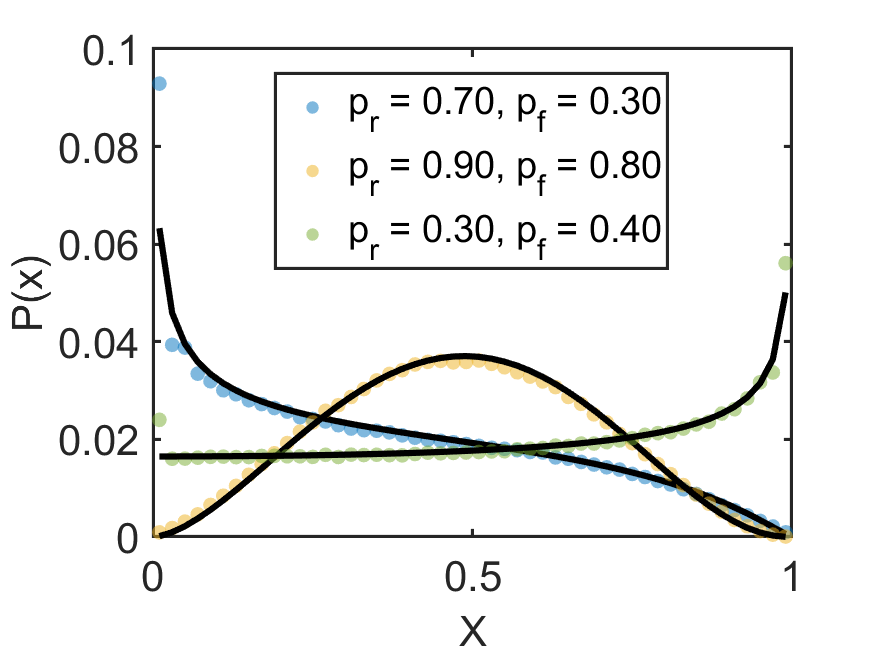}}
	%\vspace{-3.cm}
	\caption{{\bf Density profiles. } In the left panel we present the results from a numerical experiment with seed dispersal from the mainland, where in the recruitment step seeds are picked with replacement from the seed bank. The three cases (with $J=100$, $\nu = 1/J$, $p_r$ and $p_f$ values indicated in the legend) are compared with the prediction of Eq. (\ref{eq10}) (black lines). In the right panel the process includes juvenile  migration (instead of seed migration) from the mainland, $m=2$ and seeds are picked from the bank without replacement. The results are compared with the expression (\ref{eq11}) (black lines).   \label{fig5}}
\end{figure}

In that case, we must examine the dynamics of $P(x,t)$, the probability of finding the focal species with fraction $x$ at time $t$. This quantity satisfies the Fokker-Planck equation:
\begin{equation}
    \frac{\partial P(x,t)}{\partial t} = \frac{\partial^2}{\partial x^2} \left(\frac{{\rm Var(x)}}{2}  P(x,t)\right) -\frac{\partial}{\partial x} [\mu(x) P(x,t) + \nu x P(x,t) - \nu (1-x) P(x,t)],
\end{equation}
  At equilibrium  $\partial P(x,t)/\partial t = 0$. In the case with replacement [using Eq. (\ref{variance})], one obtains
\begin{equation} \label{eq10}
    P_{eq}(x) =A \left( \frac{(1-x)^{2Jp_r\nu-1}x^{2Jp_f\nu-1}}{(p_r(1-x)+p_f x)^{2J(p_r+p_f)\nu-1}}\right),
\end{equation}
where $A$ is a normalization constant.
In the case without replacement [Eq. (\ref{var2}) for the variance],
\begin{equation} \label{eq11}
    P_{eq}(x) =A \left( \frac{(1-x)^{\frac{2J m p_r\nu}{m-p_r}-1}x^{\frac{2J m p_f \nu}{m-pf}-1}}{( m p_r(1-x)+m p_f x-p_f p_r)^{\frac{2Jm(m[p_r+p_f]-2 p_f p_r)\nu}{(m-p_f)(m-p_r)}-1}}\right),
\end{equation}
Figure \ref{fig5} shows the agreement between these formulas and the numerical experiment.

In general, there is a sharp transition at $p_f = 2 / J \nu$. If $p_f < 2 / J \nu$, the distribution peaks in the vicinity of $x = 0$, indicating a high probability of the system spending time close to focal extinction. Conversely, if $p_f > 2 / J \nu$, the pdf goes to zero at $x = 0$, reflecting the rarity of the system approaching focal extinction. The same behavior is observed in the vicinity of $x = 1$, below and above $p_r = 2 / J \nu$.

\section{Selection} \label{sec6}

So far, we have considered dynamics that are, on average, neutral, meaning that each individual—whether focal or rival—produces the same average number of seeds, with differences between the types arising solely from stochasticity. Broadly speaking, we observed that the more stable species (the one experiencing less stochasticity) gains a competitive advantage.

In this section, we will explore what happens when one species produces more seeds on average than the other, representing the classic case of selection. The most intriguing scenario emerges when the species favored by stochasticity (i.e., the more stable one) is disadvantaged in terms of selection, meaning it produces fewer seeds.

To simplify the analysis, we will focus on a single property: the probability of ultimate fixation, $\Pi$. Additionally, for simplicity, we assume $p_f = 1$. We believe that the insights gained from this simplified analysis, particularly regarding the interplay between the advantage at the level of stochasticity and the deterministic advantage, are generic.

Therefore, we analyze the model described above with one key difference: each individual of the focal species produces, on average, $m e^s$ seeds, while each individual of the rival species produces $m$ seeds. An analysis of the process, as described earlier, reveals that the change in variance is negligible, whereas the mean changes according to:

\begin{equation} \label{mean1}
\mu =  \overline {\Delta x} = (n_{t+1}-n_t)/J = \frac{x(1-x)(p_f-p_r)}{Jp_fp_r}+sx(1-x).
\end{equation}

Therefore, the chance of ultimate fixation becomes:
\begin{equation} \label{eq16}
    \Pi = \frac{1-p_r^\gamma(p_r(1-x)-x)^{1-\delta}}{1-p_r^\gamma}
\end{equation}
where
\begin{equation}
\gamma = \frac{1+p_r (2Js-1)}{1-p_r} \qquad \delta = 2\frac{1+p_r (Js-1)}{1-p_r}.
\end{equation}

\begin{figure}[h]
	%\vspace{-3.cm}
	\centering{
\includegraphics[width=8cm]{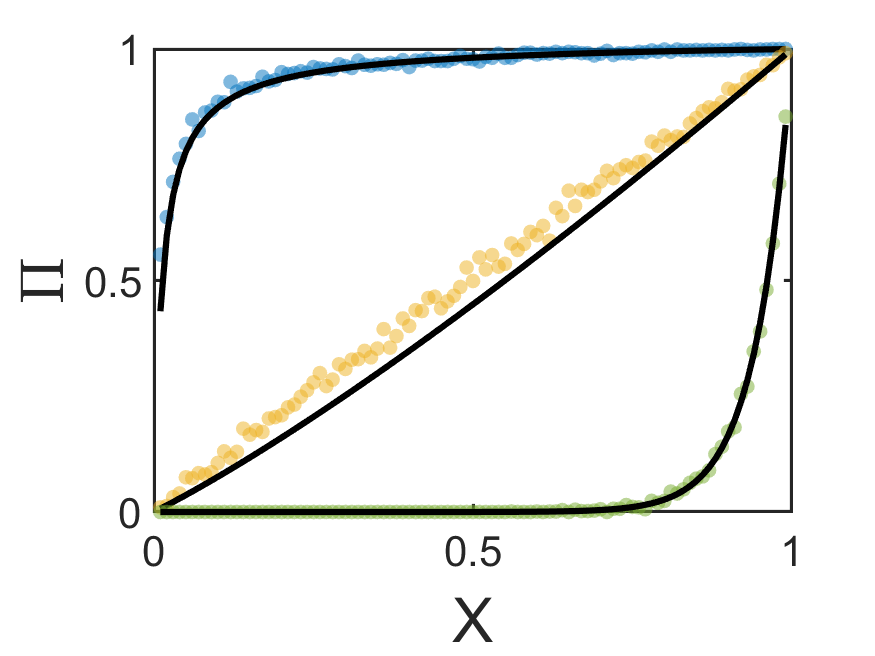}}
	%\vspace{-3.cm}
	\caption{{\bf Effect of selection.}  The chance of ultimate fixation, $\Pi(x)$, as measured in a numerical experiment with $J=100$, $s = -0.1$ and $p_f=1$,  is plotted for $p_r=0.01$ (blue), $p_r=0.1$ (yellow) and $p_r = 0.9$ (green). Seeds are picked with replacement, so the black lines represent the theoretical prediction of Eq. (\ref{eq16}). Note that the $1/J|s| = 0.1$, so we are in the strong selection regime. Selection acts against the focal species, so if $p_r$ is close to one (meaning that stochasticity advantage of the focal species is weak) the chance of the focal to win is tiny unless its initial fraction is around $x=1$ (green). As $p_r$ decreases, the stochasticity advantage of the focal dominates the selective advantage of the rival and the roles are reversed. The transition occurs at $p_r = 1/J|s|$.   \label{fig6}}
\end{figure}

The excellent fit between this expression and the numerical results is demonstrated in Figure \ref{fig6}.

The expression (\ref{eq16}) converges to (\ref{fix1}) in the limit of no selection, namely, $s \to 0$ (and $p_f =1$). When $p_r =1$ (no difference in the strength of demographic noise) Eq. (\ref{eq16})  converges to the correct expression for $\Pi$ under such circumstances~\cite{karlin1981second},
\begin{equation} \label{fix3}
    \Pi = \frac{1-e^{-2sJx}}{1-e^{-2sJ}}.
\end{equation}

Interestingly, in the strong selection limit, i.e.,  when $Js \gg 1$, $\gamma = \delta = 2Jsp_r/(1-p_r)$. In that case the chance of ultimate fixation takes the form
\begin{equation}
    \Pi = \frac{1-e^{-2sJ\Theta(x)}}{1-e^{-2sJ\Theta(1)}}.
\end{equation}
where
\begin{equation}
    \Theta(x) = \frac{p_r \ln (1-x+x/p_r)}{1-p_r}.
\end{equation}
At $p_r \to 1$, $\Theta(x) \to x$, in agreement with (\ref{fix3}). At any other $p_r <1$,  $\Theta(x) \approx x(1+x(1-p_r)/p_r $ at $x \ll 1$.  Therefore, $\Theta$ provides appreciable corrections to the $p_r=1$ predictions only above $x_c = p_r/(1-p_r)$. However, when $x>1/Js$, $\Pi(x) \approx 1$. As a result, $p_r$ yields corrections to $\Pi$ only when $p_r<1/J|s|$ (This point was highlighted by \citet{jops2023life}.). In the strong selection regime, this implies that there is no effect of $p_r$ until it is taken to very small values.

\section{environmental stochasticity} \label{sec7}

Now we aim to analyze the impact of environmental stochasticity with differing intensities for the focal species and its rival. Since we have observed that stochasticity reduces the effective fitness of the more stochastic species, we will examine a case where the rival is more affected by demographic stochasticity than the focal, while the focal is more affected by environmental stochasticity than the rival.

Therefore, we will consider a scenario where the dynamics of the rival are as described above. For the focal species, we assume $p_f = 1$ (i.e., there is no demographic stochasticity during the seed production stage), but the number of seeds per individual is not $m$, as it was previously, but rather $m(1 + \eta)$, where $\eta$ is chosen randomly each year. Consequently, the number of seeds will be:

\begin{equation} \label{eq30}
    S_f = m n (1 \pm \eta) \qquad S_r = m(J-n) \pm m\sqrt{(J-n)p(1-p)}.
\end{equation}
We have taken $\eta$ as a normal deviate with zero mean and variance $\sigma^2$.

To implement the diffusion approximation, we need the mean and the variance of $\Delta x$ per year. The same calculations described above yield, in our case, quite complicated expressions. Assuming weak environmental variations,  we have focused on the leading correction in $\sigma$. In this limit the  mean and the variance are,
\begin{equation}
    \mu = \frac{(1-p)x(1-x)}{Jp} - \sigma^2 x^2 (1-x),
\end{equation}
and
\begin{equation}
    {\rm Var}[x] = \frac{[p(1-x)+x]x(1-x)}{Jp} + \sigma^2 x^2 (1-x)^2.
\end{equation}
As can be seen from the expression for the mean, stochasticity harms competitive ability. Therefore, demographic stochasticity reduces the fitness of the rival species (an effect that disappears in the limit $p_r = 1$), while environmental stochasticity harms the focal species.

Another observation is that, to this order of approximation, the results depend only on two parameters: $p_r$ and the combination
\begin{equation}
   \kappa \equiv Js^2
\end{equation}

A tedious but elementary calculation shows that the chance of ultimate fixation takes the form,
\begin{equation} \label{eq21}
    \Pi = \frac{1-[F(x)]^\rho}{1-[F(1)]^\rho}
\end{equation}
where
\begin{eqnarray}
  \rho &=& \frac{p+\kappa p -1}{\sqrt{1+2p(\kappa-1)+(p+p\kappa)^2}} \nonumber \\  F(x) &=& \frac{p (2 + x (-1 + \kappa)) +
 x (1 + \sqrt{1 + 2 p (-1 + \kappa] + (p + p \kappa)^2]}}{x + p (2 + x (-1 + \kappa])) -
 x \sqrt{1 + 2 p (-1 + \kappa]) + (p + p \kappa])^2}}.
\end{eqnarray}

\begin{figure}[h]
	%\vspace{-3.cm}
	\centering{
\includegraphics[width=8cm]{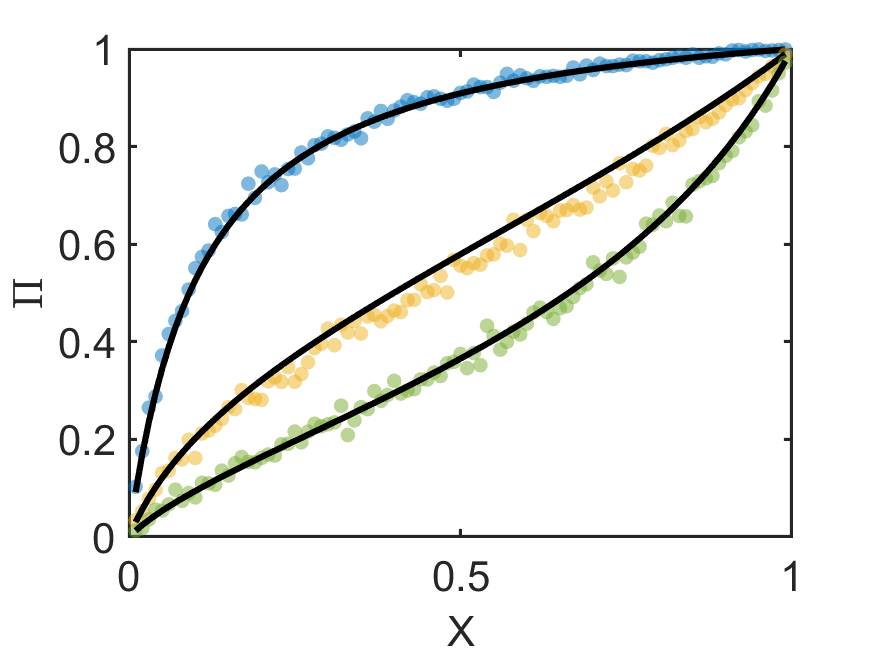}}
	%\vspace{-3.cm}
\caption{{\bf Effect of Environmental Stochasticity.}
The probability of ultimate fixation, $\Pi(x)$, measured in a numerical experiment with $J = 100$, $p_r = 0.1$, and $p_f = 1$, is plotted for $\sigma = 0.032$ (blue), $\sigma = 0.4$ (yellow), and $\sigma = 0.575$ (green). The corresponding values of $\kappa$ are $0.01$, $16$, and $32$. Seeds are picked with replacement. The black lines show the theoretical predictions of Eq. (\ref{eq21}). In the small-$\kappa$ regime, the focal species is advantageous because the rival suffers from additional demographic stochasticity. Conversely, in the large-$\kappa$ regime, environmental stochasticity dominates, rendering the focal species weaker than its rival.
\label{fig7}}
\end{figure}

The validity of Eq. (\ref{eq21}) is demonstrated in Fig. \ref{fig7}. Overall, we observe that as $J \to \infty$, while all other parameters are held constant, $\kappa$ diverges, making environmental stochasticity the dominant process. Differences in demographic noise are expected to govern effective fitness differences primarily in relatively small communities.

\section{Stochasticity and the effective fitness }

Why does a higher level of stochasticity, whether demographic or environmental, harm a species' chances of winning in competition against other species? To understand this, it is essential to consider the dynamics of the log population size, which is the decisive factor in determining effective fitness~\cite{kelly1956new,lewontin1969population}. For stochasticity with a mean of zero, such as demographic stochasticity ($n \to n \pm \sqrt{n}$), or environmental stochasticity of the type described in Equation (\ref{eq30}) ($n \to n(1+\eta)$), the geometric mean is negative. In other words, higher stochasticity creates an average drift towards extinction in the log population size space. Consequently, species exposed to higher levels of stochasticity (demographic or environmental) will exhibit lower effective fitness.

To be more precise and to account for the dynamics of a rival species, consider the model described in the previous section, now without demographic stochasticity (adults, seed and juvenile numbers are non-integers). Each individual of the focal species produces $1+\eta$ seeds, compared to each seed produced by an individual of the rival species. Here, $\eta$, as before, is a random variable with a mean of zero and variance $\sigma^2$. The seed bank ratio is thus given by:
\begin{equation}
x_{t+1} = \frac{x(1+\eta)}{x(1+\eta)+(1-x)} = \frac{x(1+\eta)}{1+ \eta x}.
\end{equation}

Therefore, the {\it logit} parameter, defined as,
\begin{equation}
    z \equiv \ln \left(\frac{x}{1-x}\right),
\end{equation}
satisfies
\begin{equation}
    z_{t+1} = z +\ln (1+\eta).
\end{equation}
If $\eta \ll 1$, the mean over $\ln (1+\eta)$ is approximately $-\sigma^2/2$. Thus, balanced stochasticity (with a mean of $\eta$ equal to zero) generates a consistent leftward bias in the logit space.

This analysis suggests that when stochasticity is balanced in log space, the more stochastic species will not inherently be at a disadvantage compared to the stable species. In a model with no demographic stochasticity, this corresponds to each focal individual producing $e^{\eta}$ seeds, yielding:
\begin{equation}
x_{t+1} = \frac{x e^{\eta}}{x e^{\eta} +(1-x)},
\end{equation}
and:
\begin{equation}
z_{t+1} = z +\eta.
\end{equation}

Here, populations undergo an unbiased random walk in logit space, with no inherent preference toward one species, as analyzed in detail by \cite{dean2020stochasticity,pande2022quantifying}. Therefore, log-balanced stochasticity does not inherently favor the more stable species.

However, we emphasize that interactions between environmental and demographic stochasticity can still produce weak disadvantages for the more fluctuating species. This effect is illustrated in Figure 8, showing fixation probabilities from a numerical experiment where:
\begin{equation} \label{eq31}
S_f = m n e^{\eta} \qquad S_r = m(J-n) \pm m\sqrt{(J-n)p(1-p)}.
\end{equation}

While environmental stochasticity has a weaker effect, it still slightly reduces the focal species' probability of winning the competition.

\begin{figure}[h]
\centering{
\includegraphics[width=8cm]{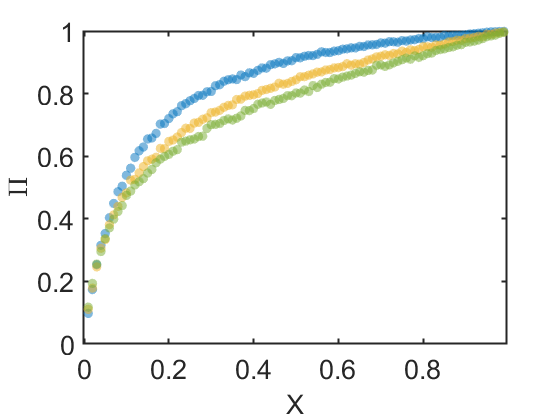}}
\caption{{\bf Effect of Log-Space-Balanced Stochasticity.} Results are presented for the same numerical experiment as in Fig (\ref{fig7}), with the only difference being that the environmental stochasticity satisfies $S_f = m n e^{\eta}$. While the effect of environmental stochasticity on the effective fitness of the focal species remains negative, it is notably weaker.
\label{fig8}}
\end{figure}

\section{Discussion}

The results presented above reveal several general insights:

First, we observed no interference between $p_r$ and $p_f$. Given one of these variables, the total lifetime of the system ($T_A$ or $T_F$) decreases monotonically as the other variable approaches zero, meaning as the stochasticity of the second species increases. In equilibrium states, such as those described in the presence of migration or mutations, the probability of finding a particular species near extinction depends independently on $p_f$ for the focal species and $p_r$ for its rival.

Second, the strength of demographic stochasticity is a decisive factor primarily in small populations. In smaller groups, random fluctuations in individual reproductive success can have a decisive impact on population dynamics. In contrast, in larger populations, these individual-level random events average out, and their relative importance diminishes. Instead, fitness differences or environmental stochasticity become the dominant drivers of population dynamics. This underscores the scale-dependent nature of stochastic effects and the necessity of accounting for population size when interpreting ecological outcomes.

Despite these insights, many questions remain open, both from empirical and theoretical perspectives.

From a theoretical perspective, achieving a comprehensive understanding of key quantities such as fixation probability, exit times, and equilibrium states is essential, particularly in scenarios where fitness differences, environmental stochasticity, and demographic stochasticity all interact. These interactions can produce non-intuitive outcomes, making it challenging to predict long-term population dynamics. Furthermore, extending these analyses to systems with multiple species and intricate community structures remains a significant theoretical hurdle. A key question is how variations in stochasticity levels across species translate into differences in species richness, abundance distributions, and overall community evenness.

On the empirical side, one key question is whether population matrix models can reliably estimate differences in environmental stochasticity between species. For instance, if environmental factors such as temperature or rainfall are known to affect seed germination probabilities, species with critical life stages tied to germination might experience greater impacts from environmental stochasticity, potentially reducing their effective fitness.

Another important empirical question concerns determining relevant population sizes. Estimating effective population sizes, for example, through neutral mutation markers, could offer valuable insights into the relative roles of stochasticity differences versus deterministic fitness differences in shaping population outcomes.

Our understanding of the mechanisms governing ecological dynamics remains incomplete. However, given the ecological and conservation significance of these processes, each incremental step forward can have profound implications. The study by \citet{jops2023life} introduced a novel and valuable perspective on species coexistence under stochastic dynamics. We hope our findings provide clarity on fundamental aspects of this problem and serve as a foundation for future research aimed at addressing these unresolved questions.

{\bf Acknowledgments:} N.M.S acknowledges support from Israel Ministry of Science (Italy-Israel cooperation, grant no. 7578).
\bibliography{ref}

\end{document}